\documentclass[twocolumn,preprintnumbers,amsmath,amssymb,showpacs,aps,superscriptaddress]{revtex4}
\usepackage{graphicx}
\usepackage{dcolumn}
\usepackage{bm}
\usepackage{amsmath}
\usepackage{amssymb}
\usepackage{epsf}
\usepackage{epstopdf}
\usepackage{multirow}
\usepackage{color}
\definecolor{Red}{rgb}{1.0,0,0}
\definecolor{Blue}{rgb}{0,0,1.0}
\newcommand{\be}{\begin{equation}}
\newcommand{\ee}{\end{equation}}
\newcommand{\bea}{\begin{eqnarray}}
\newcommand{\eea}{\end{eqnarray}}

\newcommand{\upstate}{{|\!\!\uparrow\rangle}}
\newcommand{\downstate}{{|\!\!\downarrow\rangle}}
\newcommand{\allspinsup}{{|\!\!\uparrow\uparrow...\uparrow\rangle}}

\begin{document}
\title{Spin Dephasing as a Probe of Mode Temperature, Motional State Distributions, and Heating Rates in a 2D Ion Crystal}
\author{Brian C. Sawyer}
\email{brian.sawyer@boulder.nist.gov}
\author{Joseph W. Britton}
%\author{Carson A. Teale}
%\affiliation{Time and Frequency Division, National Institute of Standards and Technology, Boulder, CO  80305}
%\author{Adam C. Keith}
%\altaffiliation{Department of Physics, North Carolina State University, Raleigh, NC 27695}
%\author{C.-C. Joseph Wang}
%\author{James K. Freericks}
%\affiliation{Department of Physics, Georgetown University, Washington, DC 20057}
%\author{Hermann Uys}
%\affiliation{Council for Scientific and Industrial Research, Pretoria, South Africa}
%\author{Michael J. Biercuk}
%\affiliation{Centre for Engineering Quantum Systems, School of Physics, The University of Sydney, NSW Australia}
\author{John J. Bollinger}
\affiliation{Time and Frequency Division, National Institute of Standards and Technology, Boulder, CO  80305}
%\date{}

%Abstract restricted to 600 characters including spaces.
\begin{abstract}
We employ spin-dependent optical dipole forces to characterize the transverse center-of-mass (COM) motional mode of a two-dimensional Wigner crystal of hundreds of $^9$Be$^+$. By comparing the measured spin dephasing produced by the spin-dependent force with the predictions of a semiclassical dephasing model, we obtain absolute mode temperatures in excellent agreement with both the Doppler laser cooling limit and measurements obtained from a previously published technique (B. C. Sawyer et al. Phys. Rev. Lett. \textbf{108}, 213003 (2012)). Furthermore, the structure of the dephasing histograms allows for discrimination between initial thermal and coherent states of motion. We also apply the techniques discussed here to measure, for the first time, the ambient heating rate of the COM mode of a 2D Coulomb crystal in a Penning trap.  This measurement places an upper limit on the anomalous single-ion heating rate due to electric field noise from the trap electrode surfaces of $\frac{d\bar{n}}{dt}\sim 5$ s$^{-1}$ for our trap at a frequency of 795 kHz, where $\bar{n}$ is the mean occupation of quantized COM motion in the axial harmonic well.
\end{abstract}

\pacs{52.27.Jt, 52.27.Aj, 03.65.Ud, 03.67.Bg}
%52.27.Jt - Plasmas: non-neutral
%52.27.Aj - Plasmas: single-component
%03.65.Ud - Quantum Entanglement
%03.67.Bg - Quantum Information: Entanglement Production

%PRL - Body (figs + captions + text + equations - Acknowledgements - Refs - Author - title) limited to 3500 words
\maketitle

\section{Introduction}
Laser-cooled ions stored in radiofrequency (RF) or Penning traps readily form crystalline arrays. Sensitive measurements of the motion of ions in these arrays are important for a variety of studies in atomic physics, quantum information science, and plasma physics. In atomic physics the small residual motion of trapped ions produces systematic errors that can limit the performance of atomic clocks and precision measurements~\cite{Rosenband08,Shiga11}. In quantum information science, where trapped-ion crystals provide a promising platform for quantum computation and simulation~\cite{Blatt08,HomeAMOAdvances13,Blatt12,Islam13,Britton12}, the residual motion of the trapped ions produces infidelities that require careful evaluation.  In plasma physics, trapped-ion crystals provide a convenient laboratory platform for studies of strongly coupled plasmas, which model dense astrophysical matter~\cite{vanHorn91,Ichimaru87,Dubin94,Baiko09}.  Careful measurements of ion motion are used to determine the ion energy and the plasma coupling. Energy transport studies require detailed measurements of the ion motion, typically as a function of time and resolved spatially or between different modes~\cite{Anderegg09}.

Here we discuss a new technique for measuring the temperature and, more generally, the energy state distribution of a trapped-ion crystal. The technique is mode specific in that it can resolve the energy distribution of different modes of the crystal. We demonstrate the technique by presenting measurements of the energy distribution of the axial center-of-mass (COM) mode of a single-plane array of several hundred Be$^{+}$ ions stored in a Penning trap. The technique requires isolating and controlling a two-level system -- an effective spin-1/2 in each ion -- and employs a weak, global spin-dependent force that couples the spin and motional degrees of freedom of each ion. This general technique should be applicable to other systems such as neutral atoms in optical lattices~\cite{Bloch05} or opto-mechanical systems~\cite{Brahms12} where spin degrees of freedom can be controlled and coupled to motional degrees of freedom.

Spin-dependent forces are a key tool in trapped-ion quantum simulation and quantum computing work. Application of a spin-dependent force to a superposition of different spin states can generate entanglement between the spins, while the concomitant coupling of the spin and motional degrees of freedom typically produces infidelities that must be mitigated~\cite{GarciaRipoll05,Kim09,Sorensen99,Sorensen00,Leibfried03,Sawyer12,Wang12}.  Here we focus on the coupling and potential entanglement between the spin and motional degrees of freedom produced by a spin-dependent force, and work in a regime where the induced spin-spin entanglement is negligible.  However, the discussion and measurements presented here provide insight into the size and nature of trapped-ion quantum gate errors produced by coupling of the spins to thermal fluctuations of the motional modes~\cite{Kirchmair09}. Spin-echo as well as other dynamical decoupling techniques can remove the coupling of the spin and motional degrees of freedom~\cite{GarciaRipoll05,Hayes11,Hayes12}, but their efficacy depends on the size of the error and the coherence of the motional state throughout an experiment, which can be evaluated with the techniques discussed here.

This study extends the results of Ref.~\cite{Sawyer12}, where we measured the decrease in the composite Bloch vector length produced by the application of a homogeneous spin-dependent optical dipole force.  We showed that this decrease (or decoherence) of the Bloch vector depended on the average energy or temperature of the initial motional state. Here we show that the dephasing responsible for this decoherence may be directly measured, revealing more detailed information about the motional state. In addition to the average energy or temperature of a mode, information on the energy distribution can also be obtained. Spin-dephasing produced through the application of a spin-dependent force provides an alternative to the well known Raman sideband technique for determining the energy distribution of motional states of trapped-ion crystals~\cite{Leibfried96}. The spin-dephasing technique is particularly well-suited for many-ion crystals, and for some setups -- in particular for higher frequency two-level systems such as the 124 GHz spin-flip transition discussed here (see Sec. II) -- can be simpler to implement.

To illustrate the basic idea of spin dephasing produced through the application of a spin-dependent force, we consider the simple case of a single trapped ion whose motional degree of freedom along the \textit{z}-axis (trap frequency $\omega_{z}$) is coupled to two internal spin states through a sinusoidally time-varying spin-dependent force. The interaction Hamiltonian for this system is
\begin{equation} \label{classical}
\hat{H}=F_{0}\cos\left(\mu t\right)\hat{z}\hat{\sigma}^{z},
\end{equation}
where $\hat{z}$ is the position operator of the ion in the \textit{z}-direction, $\hat{\sigma}^{z}$ is the Pauli spin matrix associated with the two internal energy levels, and $\mu$ is the frequency of the applied spin-dependent force. We assume the ion spin state is initialized in an equal superposition $\left\{ \left|\uparrow\right\rangle +\left|\downarrow\right\rangle \right\} /\sqrt{2}$ of the $\left|\uparrow\right\rangle ,\left|\downarrow\right\rangle $ internal levels. This spin state can be represented as pointing along the $x$-axis in the rotating frame of the Bloch sphere. Suppose the ion temperature is large compared to $\hbar\omega_{z}/k_{B}$, where $\hbar$ and $k_{B}$ are the Planck and Boltzmann constants, and we may treat the ion motion as classical. The initial motional state of the ion can then be written as $z(t)=Z_{A}\cos(\omega_{z}t+\phi)$, where $Z_{A}$ and $\phi$ fluctuate from one shot (or realization) of the experiment to the next, consistent with a thermal distribution. Application of Eq.~\ref{classical} produces an additional spin-dependent motion, but we assume this driven spin-dependent motion is small compared with the initial thermal fluctuation ($Z_{A})$ (valid for our work with hundreds of trapped Be$^{+}$ ions), and we can approximate the Hamiltonian, $\hat{H}$, as
\begin{equation}
\begin{array}{ccc}
\hat{H} & \approx & F_{0}\cos\left(\mu t\right)Z_{A}\cos\left(\omega_{z}t+\phi\right)\hat{\sigma}^{z}\\
 & = & \frac{F_{0}Z_{A}}{2}\left\{ \cos\left[\left(\mu-\omega_{z}\right)t-\phi\right]+\cos\left[\left(\mu+\omega_{z}\right)t+\phi\right]\right\} \hat{\sigma}^{z}\:.
\end{array}
\end{equation}
For $\mu=\omega_{z}$ this Hamiltonian is simply a constant shift $F_{0}Z_{A}\cos\left(\phi\right)/\hbar$ in the frequency difference between the $\left|\uparrow\right\rangle ,\left|\downarrow\right\rangle $ levels, plus a rapidly varying term that averages to zero for time intervals long compared to $\pi/\omega_{z}$. If the spin-dependent force is applied for a time interval $\tau$, then the Bloch vector undergoes a precession by an angle $\Phi_{p}=\left(F_{0}Z_{A}\cos\left(\phi\right)/\hbar\right)\tau$. Fluctuations in $Z_{A}$ and $\phi$ from one shot (or realization) of the experiment to the next produce spin dephasing when averaged over many experimental realizations. By measuring this dephasing directly we show that it is possible to acquire information on the initial motional state (for example, the energy distribution) of the trapped-ion harmonic oscillator. The sensitivity to motion of this technique is very high. For the modest parameters used in the measurements of Sec. III and IV, $F_{0}=10^{-23}$ N and $\tau=1$ ms give $\Phi_{P}=30^{\circ}$ for $Z_{A}\simeq6$ nm. With $N\gtrsim100$ trapped ions, a $30^{\circ}$ precession is much larger than the quantum projection noise~\cite{Itano93}, and can be measured with good signal-to-noise in one experimental shot.

The rest of the manuscript is structured as follows. In Section II we describe the Penning trap setup where we implement the spin dephasing technique to characterize the energy distribution of the axial COM mode of a 2D trapped-ion crystal of hundreds of $^{9}$Be$^{+}$ ions. In Section III we discuss a more detailed dephasing model assuming a thermal distribution of coherent states. In Section IV we discuss dephasing measurements of the COM mode energy distribution for both thermal states and coherently-excited thermal states. We also present measurements of the heating rate of the axial COM mode. In Section V we summarize and conclude.

\section{Experimental Setup}

\begin{figure}[t]
\resizebox{8.0cm}{!}{
\includegraphics{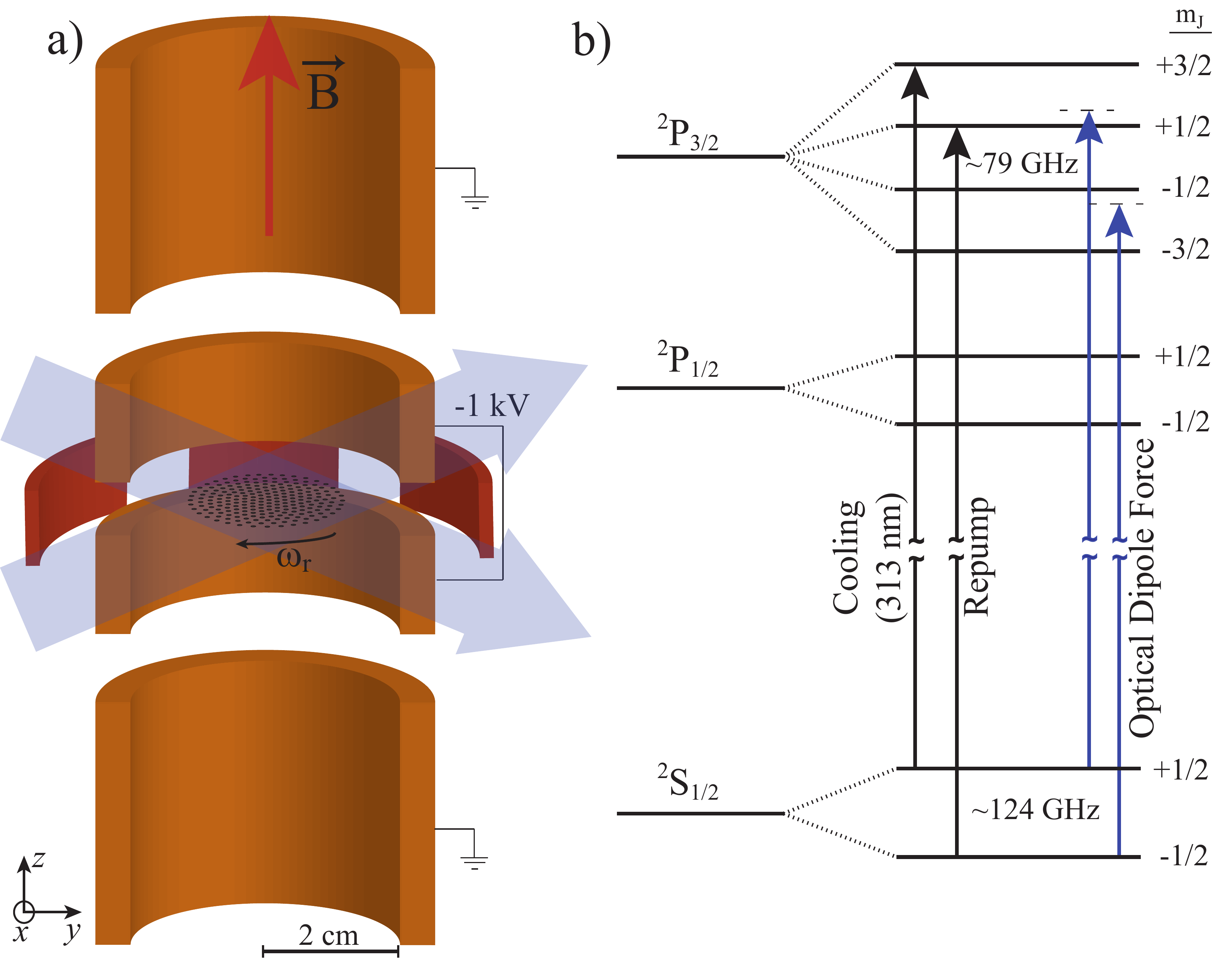}}
\caption{\label{introfig}(color online) (a) Simplified illustration of the Penning trap electrode structure in cross section (not to scale). Static voltages are applied to the electrodes in the foreground (orange), while three of the six rotating wall electrodes (red) are shown in the background. A simulated single-plane ion configuration is shown at the trap center and magnified by $\sim$100 for visibility. The 313-nm intersecting optical dipole force beams are also illustrated. (b) Low-lying electronic levels of $^9$Be$^+$ in the 4.46 T $B$-field of the Penning trap. Projection of total electronic angular momentum $(m_J)$ is given to the right of each level. Nuclear spin projections have been excluded for clarity. Relevant transitions near 313 nm are labeled.}
\end{figure}

As described in previous publications, we employ a Penning trap to confine crystals of hundreds of $^9$Be$^+$ ions~\cite{Britton12, Sawyer12}. Depicted in Fig.~\ref{introfig}a, the trap consists of a static electric ($E$) quadrupole produced from a stack of cylindrical electrodes (inner radius of 2.0 cm) placed within the room-temperature bore of a $\sim$4.46 T superconducting magnet. The orientation of this uniform magnetic ($B$) field defines the $z$-axis in our system. Harmonic axial ($z$-axis) ion confinement with a frequency of $\omega_z = 2\pi \times 795$ kHz is obtained by applying -1 kV to the central ring electrodes relative to grounded upper and lower endcap electrodes. The cylindrical axis of the trap electrodes is aligned with the uniform magnetic field. Radial ion confinement results from $\vec{E}\times \vec{B}$ induced rotation through the magnetic field. We apply a weak quadrupolar ``rotating wall" potential to precisely control the rotation frequency ($\omega_r$) and hence, radial confining force, of the ion cloud~\cite{Hasegawa05}. Neglecting the weak azimuthal dependence of the rotating wall potential, the following trap potential describes ion confinement in the Penning trap as seen in a frame rotating at $\omega_r$~\cite{Dubin99}:
\begin{eqnarray}
q\Phi_{\text{trap}}(r,z) &=& \frac{1}{2}M\omega_z^2 \left(z^2 + \beta_r r^2 \right), \\
\beta_r &\equiv& \frac{\omega_r(\Omega_c - \omega_r)}{\omega_z^2} - \frac{1}{2}, \label{beta}
\end{eqnarray}
where $M$ ($q$) is the mass (charge) of a single $^9$Be$^+$, $\Omega_c=2\pi \times 7.597$ MHz is the cyclotron frequency, and $z$ ($r$) is the axial (radial) distance from the trap center. For rotation frequencies where the radial confinement is weak relative to transverse confinement ($\beta_r\ll1$), we obtain a single ion plane. The rotation frequency at which the ion configuration transitions from two planes to one plane depends sensitively on the ion number~\cite{Mitchell98}. For most experiments, we operate with 100 to 300 ions, which necessitates $\omega_r\lesssim 2\pi \times 48 \text{ kHz}$ for single-plane conditions with $\omega_z = 2\pi \times 795$ kHz.

We use Doppler laser cooling along both the axial ($z$-axis, out-of-plane) and radial (in-plane) trap dimensions to produce the Coulomb crystal. As shown in Fig.~\ref{introfig}b, the hyperfine structure of $^9$Be$^+$ exhibits a strong Zeeman shift in the large uniform $B$-field of the Penning trap. We laser cool along the $\sim$313 nm $|J=1/2,m_J=+1/2\rangle \rightarrow |3/2,+3/2\rangle$ cycling transition between the $^2$S$_{1/2}$ and $^2$P$_{3/2}$ manifolds, where $J$ and $m_J$ are the total electronic angular momentum and its projection along the $B$-field axis, respectively. The linewidth of this cooling transition is $\Gamma \sim 2\pi \times 17.97$ MHz~\cite{NISTasd}, yielding a Doppler cooling limit of $\hbar \Gamma / 2 k_B \sim 0.43$ mK. For all experiments described here, the $^9$Be$^+$ are optically pumped to the $|I=3/2,m_I=+3/2\rangle$ nuclear spin state~\cite{Itano81}.

The two qubit states for our experiments are the $\upstate \equiv |1/2,+1/2\rangle$ and $\downstate \equiv |1/2,-1/2\rangle$ valence electron spin projections of the $^2$S$_{1/2}$ electronic ground state. The cooling and repump transitions illustrated in Fig.~\ref{introfig}b allow for efficient preparation of all $N$ trapped ions to the state $\upstate_N \equiv \allspinsup$. The splitting between qubit levels is $\sim$124 GHz, and we perform global qubit rotations via direct application of resonant millimeter-wave radiation to the ions. We typically achieve $\pi$-pulse times ($t_{\pi}$) of $\sim$70 $\mu$s. As discussed below, for the experiments described here we perform global readouts of the qubit state through state-dependent resonance fluorescence on the Doppler cooling transition.

\begin{figure}[t!]
\resizebox{7.5cm}{!}{
\includegraphics{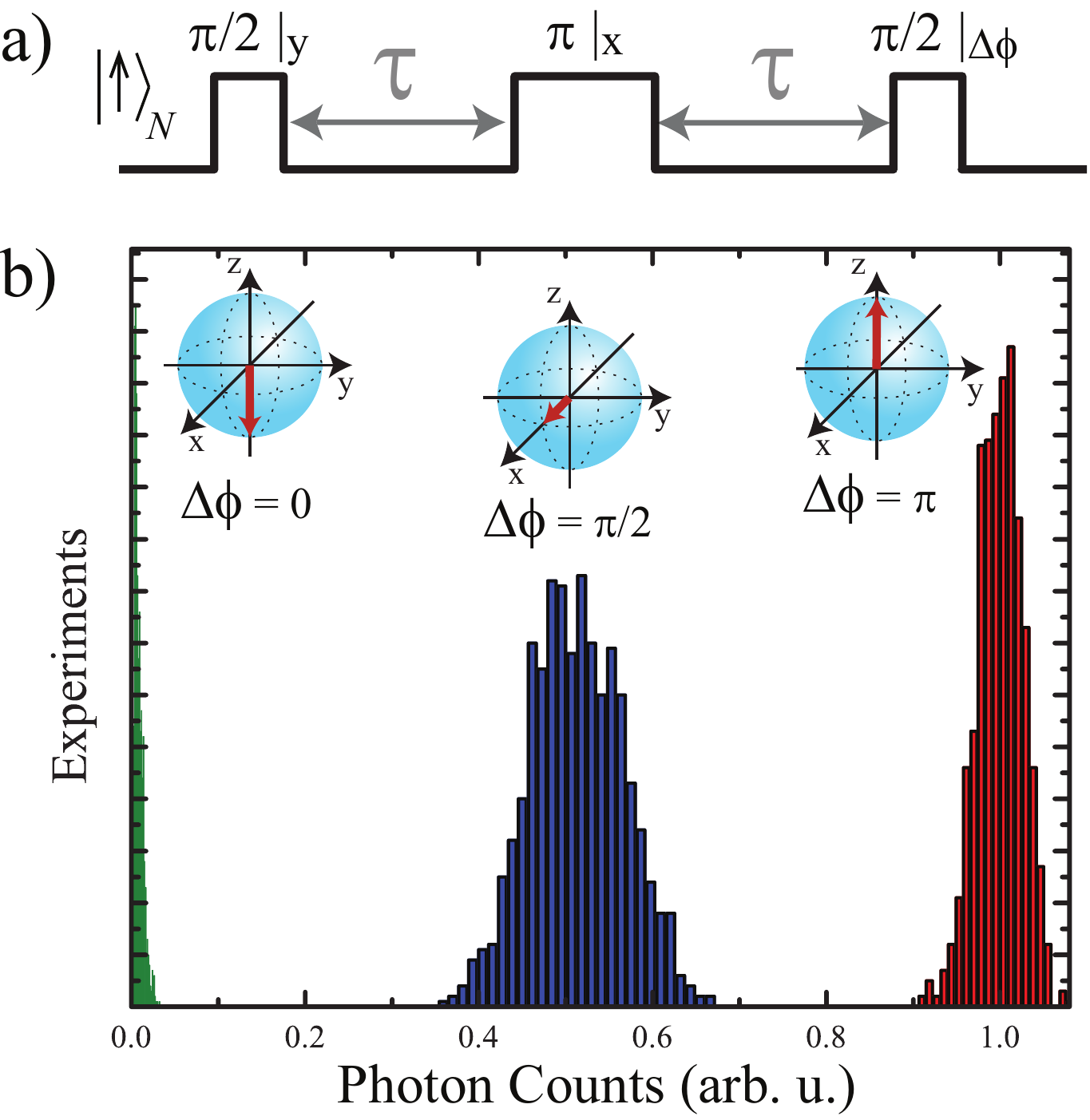}}
\caption{\label{fig1}(color online) (a) An example spin echo pulse sequence used for qubit state manipulation. Spins are initialized to $\upstate_N$, and the 124 GHz is pulsed for the given duration at a specific phase, where $\Delta\phi$ is defined relative to the $y$-axis and is varied depending on the intended final spin state. (b) Detected histograms for three collective spin orientations of the qubit ensemble consisting of $174(10)$ spins. The collective state at the end of the spin-echo sequence is represented on a Bloch sphere above the corresponding histogram. The horizontal axis is scaled such that the mean photon counts detected when in state $\upstate_N$ corresponds to unity. Each histogram is the result of 1000 state preparation and detection sequences, and bin widths have been adjusted between the three histograms to place them on the same vertical scale.}
\end{figure}

Figure~\ref{fig1}a illustrates a typical pulse sequence for qubit manipulation. We first prepare $\upstate_N$ using the Doppler cooling and repump lasers. The spin echo sequence shown includes both $\pi/2$- and $\pi$-pulses about the given Bloch sphere axes. The phase of the final pulse ($\Delta\phi$) is defined relative to that of the first $\pi/2$-pulse, which we define to be a rotation about the $y$-axis, and is varied depending on the intended final spin state. We use a spin echo sequence with free evolution periods of $\tau\sim 0.1$ to 1 ms to mitigate the deleterious effects of radial $B$-field inhomogeneity over the $\sim$400 $\mu$m ion plane diameter, and to cancel $B$-field fluctuations at frequencies below $\tau^{-1}$~\cite{Hahn50,Biercuk09nature}. After the pulse sequence, we measure the population of spins in state $\upstate$ ($P_{\uparrow}$) by switching on the Doppler cooling beams and counting scattered photons collected by an ultraviolet-sensitive photomultiplier. An $f/5$ objective imaging the side of the ion plane collects the scattered cooling photons, and a typical photon count rate per ion is $10^3$ s$^{-1}$.

The histograms of Fig.~\ref{fig1}b give experimental results for three different values of $\Delta\phi$ as measured in a system of $N=174(10)$ spins. Each color-coded histogram is the result of 1000 pulse sequences (see Fig.~\ref{fig1}a) and subsequent qubit state readouts. Bin widths for the three histograms are adjusted for clear presentation on a single vertical scale, and the horizontal axis is scaled to the photon counts collected for the state $\upstate_N$. The standard deviation of the $\Delta\phi=\pi$ histogram suggests technical noise that is comparable to shot noise with 3510 photons collected. The increase in the standard deviation of the $\Delta\phi=\frac{\pi}{2}$ data is due to quantum spin projection noise ($\propto N^{-1/2}$)~\cite{Itano93}.

We generate a spin-dependent optical dipole force (ODF) by interfering two $\sim$313 nm laser beams at the ion plane to create an optical lattice (see Fig.~\ref{introfig}a)~\cite{Sawyer12,Britton12}. The resulting lattice wavelength ($\lambda_l$) is determined by the crossing angle of the ODF beams as $\lambda_l = \lambda_{ODF}\left[2\sin(\frac{\theta_R}{2})\right]^{-1}$, where $\theta_R$ is the full beam crossing angle and $\lambda_{ODF}$ is the ODF laser wavelength. For this work, $\theta_R = 4.2(2)^{\circ}$, which results in $\lambda_l \sim 3.7$ $\mu$m. The ODF laser frequency is detuned by $\sim$20 GHz from the nearest resonances for the $\upstate$ and $\downstate$ states (see Fig.~\ref{introfig}b) and the linear polarization of each beam is chosen so as to produce a polarization gradient at the ion plane that imparts equal-magnitude, opposite-sign forces to the two qubit states with a magnitude of $\sim 10^{-23}$ N per spin at 1 W cm$^{-2}$ per beam. The optical lattice wavevector ($\overrightarrow{\Delta k}$) is oriented along the $z$-axis of the trap to preferentially excite motion transverse to the crystal plane. The two ODF laser beams are produced from a single beam using a 50/50 beamsplitter, and their relative frequency is adjusted from zero to $\sim$10 MHz using acousto-optic modulators, enabling production of a standing- or running-wave spin-dependent optical lattice.

\section{Thermal Dephasing Model}\label{Theory}
In Ref.~\cite{Sawyer12} we excited arbitrary drumhead modes of a 2D trapped-ion crystal through application of a homogeneous spin-dependent force. The spin-dependent force coupled the $^{9}$Be$^{+}$ ground-state valence electron spin and transverse motional degrees of freedom. We measured the decrease in the composite Bloch vector of the spins due to this coupling, and showed that this decrease (or decoherence) depended on the average energy or temperature of the motional state. We sketch the calculation of Ref.~\cite{Sawyer12} in Appendix~\ref{AppA}. In the Fock state basis thermal motional states are described by a diagonal density matrix. The calculation proceeds by assuming an initial Fock state $\left|n\right\rangle $ for a mode. Application of a spin-dependent force produces spin-dependent displacements of the Fock state and decoherence of the spins is naturally described in terms of spin-motion entanglement and the increasing displacement sensitivity of Fock states with $n$. Here we use a model motivated by the dephasing picture of the Introduction that does not require quantum entanglement of the spin and motional degrees of freedom for thermal excitations large compared with the ground state size.

For simplicity, we describe the dephasing model for the axial COM mode, although a generalization to other drumhead modes is straightforward. Center-of-mass motional modes play an important role in quantum information experiments with trapped ions. If all trapped ions possess the same charge-to-mass ratio, the COM mode frequency is independent of ion number and, in the case of transverse modes, constitutes the highest-frequency and longest-wavelength oscillation. In traps whose electrode dimensions are much larger than those of the ion crystal (i.e. Penning traps), the COM mode is the transverse mode most susceptible to noise from fluctuating potentials on trap electrodes.

The interaction Hamiltonian for the spins and the COM degree of freedom is
\begin{equation}\label{Hodf}
\hat{H}_{ODF}=F_{0}\cos\left(\mu t + \varphi\right)\frac{z_{0}}{\sqrt{N}}\left(\hat{a}\, e^{-i\omega_{z}t}+\hat{a}^{\dagger}\, e^{i\omega_{z}t}\right)\sum_{i=1}^{N}\hat{\sigma}_{i}^{z},
\end{equation}
where the sum is over the $N$ spins, $z_{0}=\sqrt{\hbar/\left(2M\omega_{z}\right)}$ is the ground state wavefunction size of a single trapped ion, $\hat{a}$ ($\hat{a}^{\dagger}$) are the lowering (raising) operators for the COM mode, and $\varphi$ is the ODF phase. In general the time evolution operator for the above Hamiltonian can be written as the product of a spin-dependent displacement operator,
$\exp\left(\left[\alpha\hat{a}^{\dagger}-\alpha^{*}\hat{a}\right]\sum_{i=1}^{N}\hat{\sigma}_{i}^{z}\right)$, and an evolution operator for a general Ising interaction that involves only pairwise spin interactions~\cite{Kim09,Sawyer12}. For resonant drives ($\mu\approx\omega_{z}$), the effect of the spin-dependent displacement typically dominates, and we neglect the induced Ising interaction throughout this manuscript. In this case the evolution operator for $\hat{H}_{ODF}$ separates into a product of $N$ individual spin-dependent displacement operators,
\begin{equation}\label{displacement}
\hat{D}_{\text{SD}}\left(\alpha\right) = \prod_{i=1}^N \exp\left(\left[\alpha\hat{a}^{\dagger}-\alpha^{*}\hat{a}\right]\hat{\sigma}^{z}_i\right).
\end{equation}
The displacement amplitude, $\alpha$, for resonant ($\mu=\omega_z$) spin-dependent excitation of the COM mode for a time, $\tau$, and phase, $\varphi$, is
\begin{equation}\label{alpha_resonant}
\alpha(\tau,\varphi) = -i\frac{F_0 z_0 \tau}{2 \hbar \sqrt{N}} e^{i\varphi}.
\end{equation}
For pulse sequences involving separated periods of ODF excitation (e.g. spin echo), the final motional displacement is simply a sum of individual displacements of the form of Eq.~\ref{alpha_resonant} with appropriate phases ($\varphi$) and times ($\tau$) for each of the ODF excitations within the sequence. We define spin-independent displacements, $\hat{D}(\alpha_0)$, in the usual way:
\begin{equation}
\hat{D}\left(\alpha_0\right)=\exp\left(\alpha_0\hat{a}^{\dagger}-\alpha_0^{*}\hat{a}\right).
\end{equation}

In contrast to Ref.~\cite{Sawyer12}, we consider the initial state of the COM mode for each experiment to be a coherent state $\left|\alpha_{0}\right\rangle $. We denote the expectation value of a quantum operator $\hat{O}$ at the end of an experiment as $\left\langle \hat{O}\right\rangle $. We then perform an average over a thermal distribution of expectation values which we denote as $\left\langle \left\langle \hat{O}\right\rangle \right\rangle _{th}$. More precisely, we calculate thermal averages of a function, $A(\xi)$, of the continuous variable $\xi \equiv |\alpha_0|^2$ as
\begin{equation}
\langle A(\xi) \rangle_{th} \equiv \beta \int_0^{\infty} A(\xi) e^{-\beta \xi} d\xi,
\end{equation}
where $\beta = \hbar \omega_z (k_B T)^{-1}$ for COM mode temperature $T$. For completeness, the Fock state calculations in the Appendix involve the corresponding thermal average over discrete Fock state expectation values, $A_n$, as
\begin{equation}
\langle A_n \rangle_{th} \equiv \left(1 -  e^{-\beta} \right)\sum_{n=0}^{\infty}A_n e^{-\beta n}.
\end{equation}

\subsection{Bloch Vector Length ($\Delta \phi = 0$)} \label{Bloch_length_section}
\begin{figure}[t!]
\resizebox{7.0cm}{!}{
\includegraphics{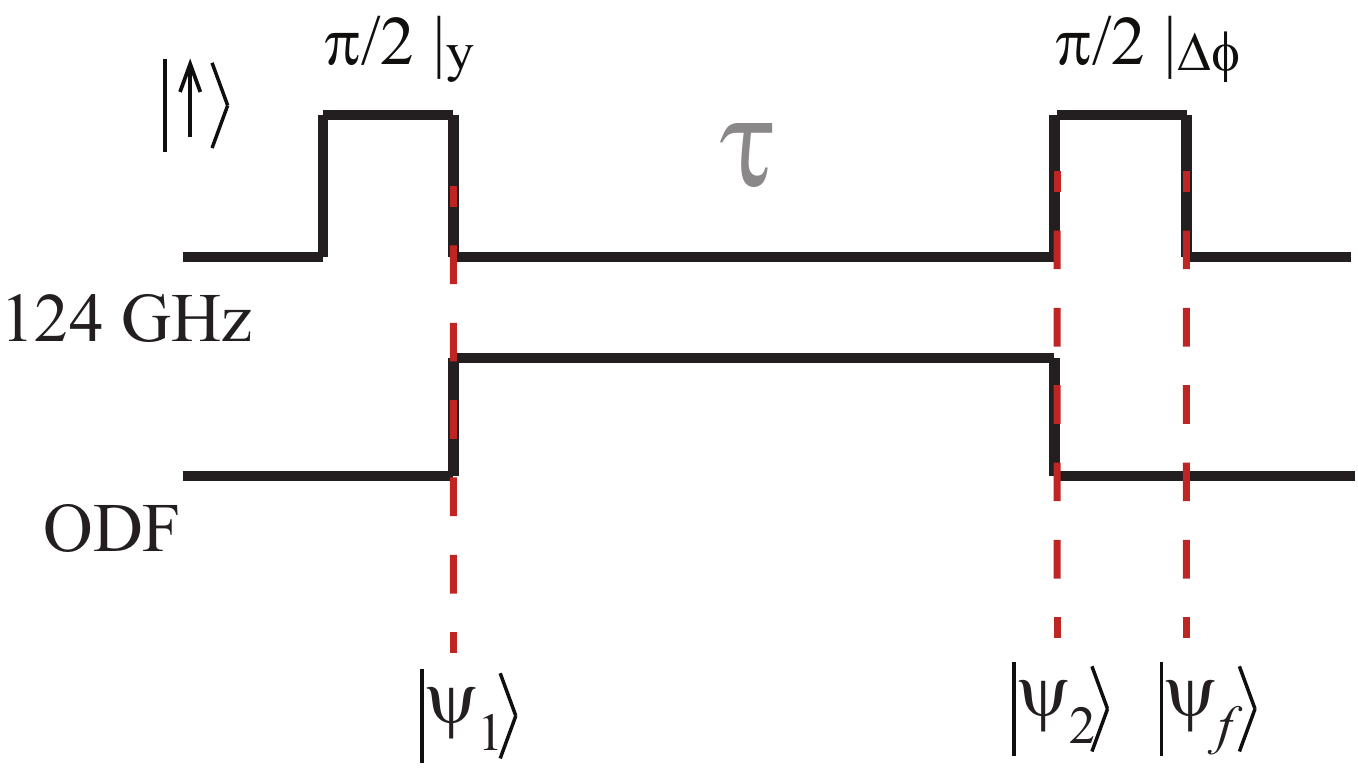}}
\caption{\label{ramtheory}(color online) Ramsey pulse sequence for global 124-GHz qubit rotations (upper) and ODF excitation (lower) as used for the derivations of Section~\ref{Theory}. The quantum states $|\psi_1\rangle$, $|\psi_2\rangle$, and $|\psi_f\rangle$ are labeled at appropriate points in the sequence as described in the text.}
\end{figure}

Here we are interested in calculating the expectation value of a component of the composite Bloch vector, $\left(\hat{S}_x,\hat{S}_y,\hat{S}_z \right)=\left(\sum_{i=1}^N \frac{\hat{\sigma}^x_i}{2},\sum_{i=1}^N \frac{\hat{\sigma}^y_i}{2},\sum_{i=1}^N \frac{\hat{\sigma}^z_i}{2}\right)$, for initial spin states which are product states. This reduces to calculating the expectation value of a component of an individual spin $\vec{\sigma}_i$. For the evolution operator of Eq.~\ref{displacement}, which is a product of commuting displacement operators involving individual spins, only the displacement operator involving $\hat{\sigma}^z_i$ non-trivially enters into the calculation. We assume that each experiment begins with the state, $\upstate |\alpha_0\rangle = \upstate \hat{D}(\alpha_0)|0\rangle$, where $|\alpha_0\rangle$ is a coherent state of COM motion. In Sec.~\ref{Bloch_length_section} and~\ref{Dephasing}, we consider a Ramsey pulse sequence consisting of two $\pi/2$ pulses separated by a time $\tau$ as shown in Fig.~\ref{ramtheory}. For this Ramsey sequence consisting of a single ODF excitation period, Eq.~\ref{alpha_resonant} is used to calculate spin-dependent displacements, $\alpha$. The measurements of Section~\ref{COMsection} involve spin echo sequences, but all of the theory results of Sections~\ref{Bloch_length_section} and~\ref{Dephasing} apply with small modifications for calculating the spin-dependent displacement, $\alpha$.

The first $\pi/2$ pulse of the sequence of Fig.~\ref{ramtheory} yields the qubit rotation
\begin{eqnarray}
|\psi_1\rangle &=& \hat{R}(\frac{\pi}{2},0)\upstate |\alpha_0\rangle \nonumber \\
&=& \frac{1}{\sqrt{2}}\left(\upstate + \downstate \right)\hat{D}(\alpha_0)|0\rangle,
\end{eqnarray}
where we define the following qubit rotation matrix
\begin{equation}
\hat{R}(\theta,\phi) = \left(%
\begin{array}{cc}
  \cos{(\frac{\theta}{2})} & -e^{-i\phi}\sin{(\frac{\theta}{2})} \\
  e^{i\phi}\sin{(\frac{\theta}{2})} & \cos{(\frac{\theta}{2})} \\
\end{array}%
\right).
\end{equation}
We now consider the effect of the spin-dependent ODF acting for the free evolution time, $\tau$. This yields the state
\begin{eqnarray}
|\psi_2\rangle &=& \frac{1}{\sqrt{2}}\upstate\hat{D}(\alpha)\hat{D}(\alpha_0)|0\rangle \nonumber \\
&& +\frac{1}{\sqrt{2}}\downstate\hat{D}(-\alpha)\hat{D}(\alpha_0)|0\rangle \\
&=& \frac{1}{\sqrt{2}} e^{-i\theta_0}\upstate |\alpha+\alpha_0\rangle \nonumber \\
&&+\frac{1}{\sqrt{2}}e^{i\theta_0}\downstate |-\alpha+\alpha_0\rangle, \label{psi2coh}
\end{eqnarray}
where $\theta_0 = \text{Im}\{\alpha\alpha_0^{\ast}\}$. It is useful to pause at Eq.~\ref{psi2coh} before applying the final microwave pulse and evaluate $\langle \hat{S}_x \rangle$ and $\langle \hat{S}_y \rangle$. We find
\begin{eqnarray}
\langle \hat{S}_x \rangle &=& \frac{N}{2}\langle \hat{\sigma}^x \rangle = \frac{N}{2}\cos{\left( 4\, \text{Im}\{\alpha^{\ast}\alpha_0\}\right)e^{-2|\alpha|^2}} \label{Sxcoh}\\
\langle \hat{S}_y \rangle &=& \frac{N}{2}\langle \hat{\sigma}^y \rangle = \frac{N}{2}\sin{\left( 4\, \text{Im}\{\alpha^{\ast}\alpha_0\}\right)e^{-2|\alpha|^2}}. \label{Sycoh}
\end{eqnarray}
From Eqs.~\ref{Sxcoh} and~\ref{Sycoh}, we see that the ODF has caused a coherent rotation of the composite Bloch vector about the $z$-axis by
\begin{equation}
\theta_{coh}=\arctan{\left(\langle \hat{S}_y\rangle/ \langle \hat{S}_x\rangle\right)}=4\, \text{Im}\{\alpha^{\ast}\alpha_0\}.
\end{equation}
The effect of spin-motion entanglement is reflected in the term $e^{-2|\alpha|^2}$ of Eqs.~\ref{Sxcoh} and~\ref{Sycoh}, which deviates negligibly from unity for the dephasing measurements discussed in Sec. IV. Thermal averages may be performed over the continuous variable $\xi \equiv |\alpha_0|^2$, where the magnitude $\xi$ is now weighted according to Boltzmann statistics and the phase of $\alpha_0 \in \mathbb{C}$ is evenly distributed over $2\pi$ radians. Defining $\alpha = |\alpha|e^{i\phi'}$ and $\alpha_0 = |\alpha_0|e^{i\phi_0}$, we calculate:
\begin{small}
\begin{eqnarray}
\langle \theta_{coh}^2 \rangle_{th} &=& \frac{\beta}{2\pi}\int_0^{2\pi} \int_{0}^{\infty} 16 |\alpha|^2 \xi\sin^2{(\phi' - \phi_0)} e^{-\beta\xi}d\xi d\phi_0 \nonumber \\
&=& 8|\alpha|^2\beta^{-1}. \label{thetaCoh}
\end{eqnarray}
\end{small}
In Sec. IV, a typical $|\alpha|$ for a 30-yN ODF driving at $\omega_z$ for 100 $\mu$s is $\sim\!0.05$, while $\beta^{-1}\sim12$ at the Be$^+$ Doppler cooling limit, producing a non-negligible rotation angle standard deviation of $\sim\!30^{\circ}$.

For completeness, we present $|\psi_f\rangle$, $P_{\uparrow}$ and $\langle \hat{S}_z \rangle$ results following the final $\pi/2$-pulse, $\hat{R}(\frac{\pi}{2},0)$,
\begin{eqnarray}
|\psi_f\rangle &=& \frac{1}{2}
\upstate \left(
e^{-i\theta_0}|\alpha+\alpha_0\rangle - e^{i\theta_0}|-\alpha+\alpha_0\rangle \right)  \nonumber \\
  &&+\frac{1}{2}\downstate \left( e^{-i\theta_0}|\alpha+\alpha_0\rangle + e^{i\theta_0}|-\alpha+\alpha_0\rangle \right) \\
P_{\uparrow}^{(\alpha_0)} &\equiv& \frac{1}{2} \left( 1 - \cos{\left(4\, \text{Im}\{\alpha^{\ast}\alpha_0\}\right)} e^{-2|\alpha|^2}\right) \\
P_{\uparrow} &=& \langle P_{\uparrow}^{(\alpha_0)} \rangle_{th}  \nonumber \\
&=& \frac{1}{2} \left( 1 - e^{-2|\alpha|^2(2\beta^{-1}+1)}\right) \label{PupCoh} \\
\langle \hat{S}_z \rangle &=& -\frac{N}{2}\cos{\left( 4\, \text{Im}\{\alpha^{\ast} \alpha_0\}\right)}e^{-2|\alpha|^2} \\
\langle \hat{S}_z \rangle_{th} &=& -\frac{N}{2}e^{-2|\alpha|^2(2\beta^{-1}+1)}.
\end{eqnarray}
We note that $\beta^{-1}=\langle |\alpha_0|^2\rangle_{th}$. Also, for $\bar{n}\gg 1$, $\beta^{-1} \sim \bar{n}$ and the $P_{\uparrow}$ of Eq.~\ref{PupCoh} agrees with the treatment that assumes a thermal distribution of Fock states (Eq.~\ref{PupFock}) given in Appendix~\ref{AppA}, but offers a more classical description. That is, for each possible coherent state amplitude and phase, the result of the experimental sequence is that the composite Bloch vector undergoes \textit{coherent} rotation by some angle $\theta_{coh}$. Over many such experimental sequences, we measure a dephasing of the composite Bloch vector associated with the angular variance, $\langle \theta_{coh}^2 \rangle_{th}$, of Eq.~\ref{thetaCoh}.

\subsection{Dephasing ($\Delta \phi= \frac{\pi}{2}$)} \label{Dephasing}
A final qubit rotation of $\hat{R}(\frac{\pi}{2},\frac{\pi}{2})$ ($\Delta \phi = \pi/2$) transforms rotations and dephasing in the $xy$-plane of the Bloch sphere to the detection ($z$) basis. Below we calculate the thermal average of the expectation value $\langle \hat{S}_z^2 \rangle$ from which a temperature determination can be obtained. In addition, we discuss the implementation of a Monte Carlo analysis (see Sec.~\ref{COMsection}) that is in excellent agreement with measurements of thermal as well as non-thermal motional state distributions with $\bar{n}\gg1$.

For the $\Delta \phi = \frac{\pi}{2}$ pulse sequence, we obtain the following expression for $\langle \hat{S}_z \rangle$ after the final $\pi/2$ pulse:
\begin{equation}
\langle \hat{S}_z \rangle = \frac{N}{2} \sin{(4 \text{Im}\{\alpha^{\ast}\alpha_0 \})} e^{-2|\alpha|^2}. \label{SzCoh}
\end{equation}
Note that the thermal average of the expression in Eq.~\ref{SzCoh} vanishes, so we instead calculate dephasing through the second moment of $\hat{S}_z$:
\begin{eqnarray}
\langle \hat{S}_z^2 \rangle &=& \frac{1}{4}\sum_{i=1}^N\langle \hat{\sigma}^z_i \hat{\sigma}^z_i \rangle + \frac{1}{4}\sum_{i \neq j}\langle \hat{\sigma}^z_i \hat{\sigma}^z_j \rangle \\
&=& \frac{N}{4} + \frac{N(N-1)}{4} \langle \hat{\sigma}^z_1 \hat{\sigma}^z_2 \rangle \label{2ndmom}
\end{eqnarray}
where $\langle \hat{\sigma}^z_1 \hat{\sigma}^z_2 \rangle$ is an expectation value involving any two non-identical spins within the ensemble. We simplify to Eq.~\ref{2ndmom} since, for COM excitation, all spins feel a force of equal magnitude and there is no differentiation between spin pairs. In evaluating the two-spin expectation, $\langle \hat{\sigma}^z_1 \hat{\sigma}^z_2 \rangle$, only the displacement operators in Eq.~\ref{displacement} involving $\hat{\sigma}^z_1$ and $\hat{\sigma}^z_2$ non-trivially enter into the calculation. We then compute the thermal average of this expectation value and obtain:
%\begin{small}
\begin{eqnarray}
\langle \langle \hat{S}_z^2 \rangle \rangle_{th} &=& \frac{N}{4} + \frac{N(N-1)}{4}\sigma^2 \label{2ndmomcohtherm} \\
\sigma^2 &\equiv& \left\langle \langle \hat{\sigma}^z_1 \hat{\sigma}^z_2 \rangle \right\rangle_{th} \\
&=& \frac{1}{2} \left( 1 - \langle \cos{\left( 8\, \text{Im}\{\alpha^{\ast} \alpha_0\}\right)} \rangle_{th} e^{-8|\alpha|^2}\right) \\
&=& \frac{1}{2} \left( 1 - e^{-8|\alpha|^2 \left( 2\beta^{-1}+1 \right)} \right). \label{2ndmomcohshort}
%&& \frac{N}{4} +\frac{N(N-1)}{4}e^{-4|\alpha|^2}\frac{\beta}{2\pi} \int_0^{\infty} \int_0^{2\pi} \sin^2{\left(4\alpha\sqrt{\xi}\sin\phi\right)}
%e^{-\beta\xi} \, d\phi d\xi \nonumber \\
\end{eqnarray}
%\end{small}
The first term of Eq.~\ref{2ndmomcohtherm} is the contribution of spin projection noise while the second is due to dephasing. The quadratic scaling of the dephasing with $N$ relative to the linear scaling of projection noise indicates that thermal dephasing can be more accurately measured with larger ion numbers. We may recast the normalized thermal dephasing portion ($\sigma^2$) in terms of the COM mean phonon number ($\bar{n}\sim\beta^{-1}$):
\begin{equation}\label{nbartherm}
\bar{n} \sim \frac{1}{16|\alpha|^2}\ln{\left[\frac{1}{1-2\sigma^2}\right] - \frac{1}{2}}.
\end{equation}
Using Eq.~\ref{nbartherm}, we can now extract the mean phonon occupation of the COM mode from measurements of spin dephasing. To determine $\sigma^2$ we subtract the calculable spin variance ($\frac{N}{4}$) from the measured $\langle \langle \hat{S}_z^2 \rangle\rangle_{th}$ and normalize by the squared length of the composite Bloch vector ($|\vec{S}|^2=\frac{N^2}{4}$). For $N>100$, the difference between $\frac{N^2}{4}$ and $\frac{N(N-1)}{4}$ is below 1 \%. In Appendix~\ref{AppB} we derive an analogous expression to Eq.~\ref{2ndmomcohshort} using Fock states that agrees in the limit $\beta^{-1} \sim \bar{n}$.

As detailed in Sec.~\ref{COMsection}, this approach to calculating spin dephasing using coherent states motivates a straightforward Monte Carlo analysis in which, for each experiment, we choose a random initial coherent state of motion and subsequently apply the experimental pulse sequences and state readout. The initial coherent state magnitudes are weighted according to a thermal distribution while the phase is random and unweighted in the range $[0,2\pi)$. Each simulation includes a Bloch vector rotation of $\theta_{coh}$ determined by the randomized initial coherent state and fixed spin-dependent displacement $\alpha$. After many such Monte Carlo runs, we bin the outcomes into simulated histograms for direct comparison with experimental histograms.

\section{Center-of-Mass Measurements}\label{COMsection}
In this Section, we describe measurements of spin dephasing for different initial states of COM motion. We show experimentally that such measurements reveal not only the effective temperature of the COM mode ($\bar{n}$)~\cite{Sawyer12}, but allow for a more detailed characterization of the initial motional state (e.g. thermal, coherent, or a mixture of the two).
\begin{figure*}[t!]
\resizebox{14.5cm}{6.0cm}{
\includegraphics{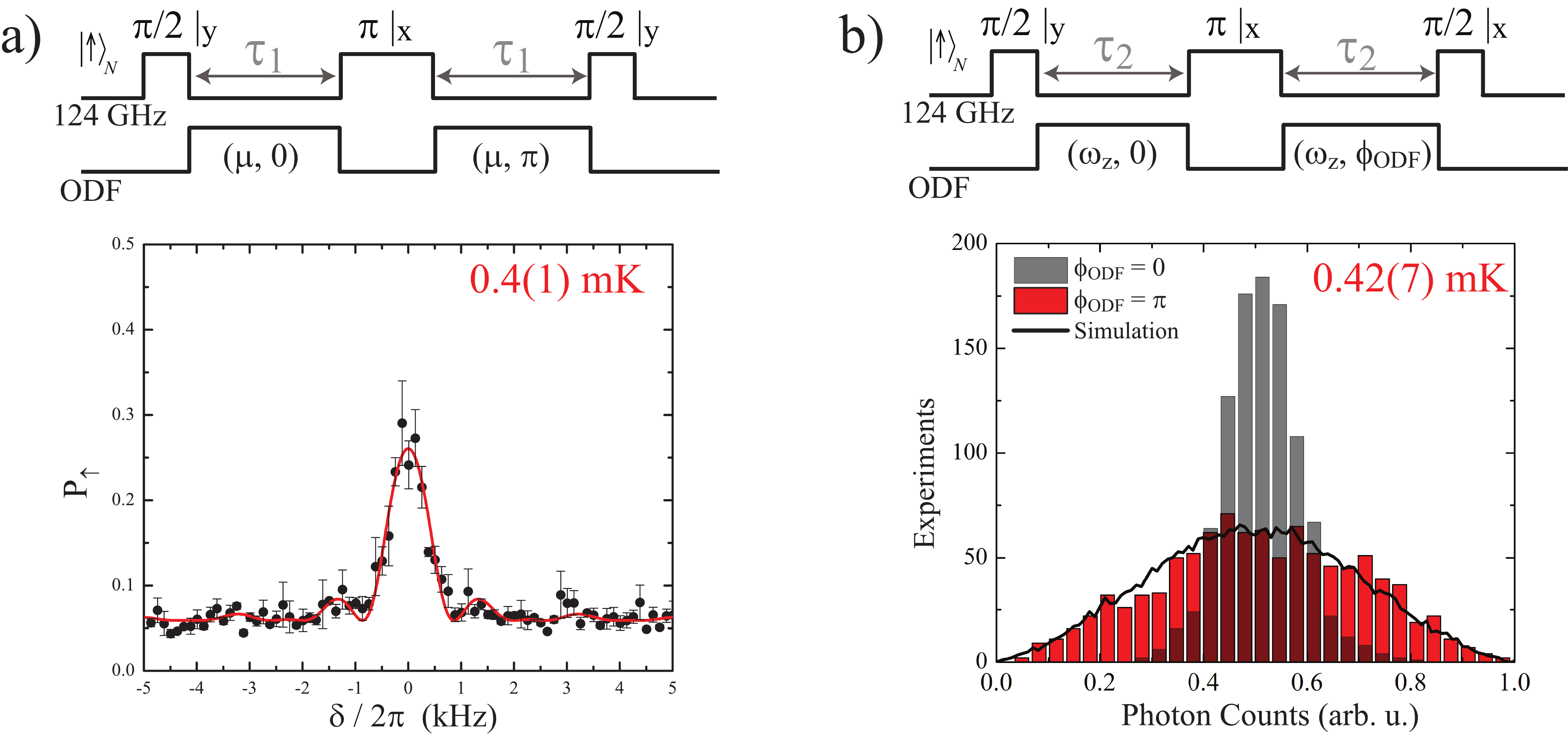}}
\caption{\label{fig2}(color online) Experimental pulse sequences for global 124-GHz qubit rotations (upper) and ODF excitation (lower) along with corresponding data. Relative phases and rotation angles are given for each qubit rotation. The ODF drive frequencies $(\mu)$ and phases $(\phi_{\text{ODF}})$ are given as $(\mu,\phi_{\text{ODF}})$. (a) Measurement of COM temperature using collective Bloch vector length. The microwave spin echo sequence produces the state $\downstate_N$ ($P_{\uparrow}=0$) in the absence of the ODF, and the ODF frequency is swept over the COM resonance ($\omega_z = 2\pi \times 795$ kHz) with $\tau_1 = 500$ $\mu$s. Frequency-dependent deviation from $P_{\uparrow}=0$ is fit to a theoretical expression derived in Ref.~\cite{Sawyer12}. The resulting experimental data (black points with error bars) is fit (solid red line) to extract a mode temperature of 0.4(1) mK given the applied ODF. (b) Extraction of COM mode temperature through direct measurement of spin dephasing. The COM mode is resonantly driven in each arm for $\tau_2=100$ $\mu$s and $\phi_{\text{ODF}}$ for the second arm is chosen to either undo ($\phi_{\text{ODF}}=0$) or enhance ($\phi_{\text{ODF}} = \pi$) the Bloch vector rotation produced by the spin-dependent force. Temperature is extracted from the excess width of the $\phi_{\text{ODF}}=\pi$ histogram (red bars) using Eq.~\ref{nbartherm}, and the solid black-line histogram is the result of a Monte Carlo simulation of 50,000 experimental runs with no adjustable parameters. We determine a COM temperature of 0.42(7) mK using this dephasing measurement.}
\end{figure*}
%
%
%A number of techniques exist for determining temperatures and heating rates in radio-frequency ion traps including Raman sideband spectroscopy and Doppler recooling~\cite{Epstein07}. Our trap magnetic field of $\sim$4.5 T and resulting qubit splitting of $\sim$124 GHz renders Raman transitions difficult to produce as they necessitate phase coherence of two 313-nm laser fields with a relative detuning near the qubit splitting -- well beyond the customary bandwidths of commercial optical modulators. Furthermore, the Doppler recooling technique is prohibitively time-expensive for low heating rates and is most useful for single-ion characterization of anomalous heating in surface traps~\cite{Epstein07}. As a result, we rely on spin decoherence from projective measurement of an entangled state of spin and motional degrees of freedom to measure transverse mode temperatures~\cite{Sawyer12}. This technique is mode-specific, applicable above or below the Doppler-cooling limit, and avoids the difficulty of producing stimulated Raman transitions in the large $B$-field of the Penning trap.

Figure~\ref{fig2}a shows the experimental pulse sequences for ODF laser beams and qubit rotations used to measure the Bloch vector length ($\Delta \phi = 0$, Sec. IIIA) as in Ref.~\cite{Sawyer12}, along with a plot of experimental data and corresponding theory fit. The collective spin of the trapped ions is first prepared in the state, $\upstate_N$, and the pulses at $\sim$124 GHz constitute a Hahn spin echo that, in the absence of the ODF beams, leaves the spins in state $\downstate_N$ with $>$99 \% fidelity~\cite{Biercuk09}. We apply the ODF laser beams during each free evolution period of the spin echo for a duration $\tau_1=500$ $\mu$s. The relative frequency of the two beams is $\mu$, and the relative phase of the ODF beat between the first and second arms is given by $\phi_\text{ODF}$. Note that, for $\phi_\text{ODF}=\pi$ and $\mu=\omega_z$, the net result of the experimental sequence is identical to that of the Ramsey sequence of Fig.~\ref{ramtheory} with no intermediate $\pi$-pulse and a single free-precession period of $2\tau_1$. This is due to the condition that the ODF on state $\upstate$ ($F_{\uparrow}$) is opposite in sign but equal in magnitude to that on $\downstate$ ($F_{\downarrow}$), where $(F_{\uparrow}-F_{\downarrow})=2 F_0$ (see Eq.~\ref{Hodf}). In this case the ODF of the second arm reinforces that of the first. In other words, the phase advance of the ODF beat by $\pi$ radians reverses the effect of the intermediate qubit $\pi$-pulse but retains the suppression of spin decoherence inherent in the spin echo. However, for the case $\mu \neq \omega_z$, the finite duration of the qubit $\pi$-pulse leads to a phase offset between the ion crystal oscillation and ODF drive at the start of the second arm given by $\delta(\tau_1 + t_{\pi})$, where $\delta \equiv (\mu-\omega_z)$, that must be included in the theoretical analysis. Assuming that we interact exclusively with the COM mode, the final position ($\alpha_{\text{SE}}$) of the ion crystal in phase space after application of the two ODF pulses of Fig.~\ref{fig2}a is given by:
\begin{eqnarray}\label{alpha}
\alpha_{\text{SE}} (\tau) &=& \frac{F_0 z_0}{2 \hbar \sqrt{N}} \frac{\left(1-e^{i\delta\tau} \right)}{\delta} \nonumber \\
&\times& \left( e^{i\varphi_0} - e^{i[\varphi_0 + \delta(\tau+t_{\pi}) + \phi_{\text{ODF}}]} \right)
\end{eqnarray}
where $\tau$ is the duration of each evolution period, $\varphi_0$ is the ODF phase at the start of the first free evolution period, and we have included the additional $\phi_{\text{ODF}}$ to denote the added phase advance in the second free evolution period of the spin echo sequence. The common phase, $\varphi_0$, does not contribute to any experimental observables and may be disregarded. The relative minus sign between phase factors in the final term of Eq.~\ref{alpha} is due to the intermediate $\pi$-pulse of the spin echo sequence, which removes displacements common to both free evolution periods since $F_{\uparrow}=-F_{\downarrow}$.

Previous experiments with ions confined within RF Paul traps have shown that Doppler laser cooling produces thermal states of ion motion~\cite{Meekhof96}. In Fig.~\ref{fig2}a, we show that measurements of the Bloch vector length under application of a spin-dependent force with $\mu \sim \omega_z$ are consistent with that of a thermal state of motion whose temperature is 0.4(1) mK, the Doppler cooling limit. The frequency width of the spectral feature is approximately given by the Fourier width of the ODF pulse duration of $2\tau_1=1$ ms, and the degree of decoherence measured near $\delta = 0$ is determined by the ODF magnitude and $\bar{n}$ according to Eqs.~\ref{PupCoh} and~\ref{alpha}~\cite{Sawyer12}. The detuning-independent background decoherence level of $P_{\uparrow} \sim 0.07$ in Fig.~\ref{fig2}a is due to spontaneous emission from the off-resonant ODF laser beams, and is fully characterized for this system~\cite{Uys10,Sawyer12}.

\subsection{Thermal Distributions}\label{thermalsection}
The measurement of Fig.~\ref{fig2}a is one of composite Bloch vector \textit{length}, and is only second-order sensitive to spin dephasing. To measure dephasing more directly and gain more complete knowledge of the spin statistics, we implement the experimental pulse sequence of Fig.~\ref{fig2}b. For this experiment, we resonantly drive the COM mode ($\delta = 0$) with the spin-dependent ODF during free-evolution periods of $\tau_2=100$ $\mu$s. We set the phase of the final $\pi/2$-pulse of the sequence to be the same as the intermediate $\pi$-pulse ($\Delta \phi = \frac{\pi}{2}$), thereby rotating any dephasing within the $xy$-plane to lie along the qubit axis. After each such sequence, we perform a projective measurement of $P_{\uparrow}$ and repeat for a total of 1000 experiments. The collection of 1000 $P_{\uparrow}$ values are binned and displayed as histograms in Fig.~\ref{fig2}b. As an additional check, we choose $\phi_{\text{ODF}}$ to be either 0 or $\pi$. In the case of $\phi_{\text{ODF}}=0$ (light gray bars), the spin echo effectively cancels the spin-dependent excitation and concomitant Bloch vector rotation, allowing for characterization of other sources of dephasing such as spin projection noise, photon shot noise, AC Stark shift fluctuations from the ODF lasers between the two spin echo arms, and excess magnetic field fluctuations not fully canceled by the spin echo. For $\phi_{\text{ODF}}=\pi$ (red bars), the precession induced by the spin-dependent force in the first arm is enhanced in the second and we observe that the detected spin variance is greatly increased relative to the $\phi_{\text{ODF}}=0$ case. We extract a COM temperature of 0.42(7) mK from this measurement by applying Eq.~\ref{nbartherm} to the excess variance of the $\phi_{\text{ODF}}=\pi$ experiments. This temperature is in excellent agreement with the theoretical Doppler cooling limit as well as the Bloch vector length measurement of Fig.~\ref{fig2}a.

To further compare the measurements of Fig.~\ref{fig2}b with the model of Sec.~\ref{Dephasing}, we use a Monte Carlo algorithm to produce simulated histograms consisting of 50,000 `detection' events. This is a factor of 50 more detections than for the experimental measurements, and is so chosen to reduce noise in the Monte Carlo histograms for clearer distinction between experiment and simulation. Importantly, the simulations have no adjustable parameters -- all inputs to the Monte Carlo algorithm are experimental parameters (e.g. $\alpha_{\text{SE}}$, $\tau$, $t_{\pi}$) or obtained from measurement (e.g. $\bar{n}$ from the measured $\sigma$). The simulated histograms are scaled vertically to match the experimental data given bin widths and total detection events.

The simulation procedure is the following: for each Monte Carlo run, $k\in \{1,...,5\times10^4\}$, we choose a random initial coherent state of COM motion given by $|\alpha_k|e^{i\phi_k}$. The probability of choosing a given magnitude, $|\alpha_k|$, is given by Boltzmann statistics for a thermal state. As such, the chosen magnitudes $|\alpha_k|$ follow a probability distribution proportional to $\exp(-\beta |\alpha_k|^2)$. The random value of $\phi_k$ is unweighted and assigned from the set $[0,2\pi)$, which assumes the phase of the initial state of COM motion is uncorrelated with that of the ODF and varies for each experimental sequence.  Following the pulse sequence of Fig.~\ref{fig2}b with $\phi_{ODF}=\pi$, we apply relevant qubit rotations in sequence with coherent $z$-axis rotations due to the spin-dependent ODF given by Eqs.~\ref{Sxcoh} and~\ref{Sycoh}, where $\alpha_k$ constitutes the initial coherent state for each run and $\alpha_{\text{SE}}$ is the spin-dependent displacement calibrated as in Refs.~\cite{Britton12,Sawyer12}. For completeness, we also include the smaller measured variance in the $\phi_{ODF}=0$ case phenomenologically by adding another, uncorrelated random rotation to the composite Bloch vector. The angle of this additional rotation follows a Gaussian probability distribution with the measured variance (light gray bars in Fig.~\ref{fig2}b)\footnote{This background dephasing contributes negligibly to the final dephasing of interest for the given experimental conditions, but is described here for completeness.}. The simulated histogram (black line) of Fig.~\ref{fig2}b shows good agreement with experimental measurements, and further supports the absolute temperature measurement of Fig.~\ref{fig2}a. Note that the higher sensitivity of the direct spin-dephasing measurements enables the use of a shorter free-evolution period. As a result, direct measurements of spin dephasing are less sensitive to COM frequency and ODF phase drift within a single experiment. The effects of spontaneous emission decoherence are also negligible in this parameter regime.

\subsection{Thermal Distributions with Large Coherent Displacements}\label{cohsection}
We now describe the spin dephasing signature of relatively large \textit{coherent} motional displacements, $\alpha_d$, acting in addition to thermal fluctuations characterized by $\bar{n}$. Such analysis may be relevant when narrow-bandwidth electric field fluctuations exist on trap electrodes at frequencies near that of the COM. As in the previous section, we assume that the phases of thermal, coherent, and ODF displacements are all uncorrelated. It may seem that little can be inferred from dephasing in the presence of coherent excitations with a random phase, but we demonstrate that a constant \textit{magnitude} is all that is required to distinguish such noise sources.

\begin{figure}[t]
\resizebox{8.7cm}{!}{
\includegraphics{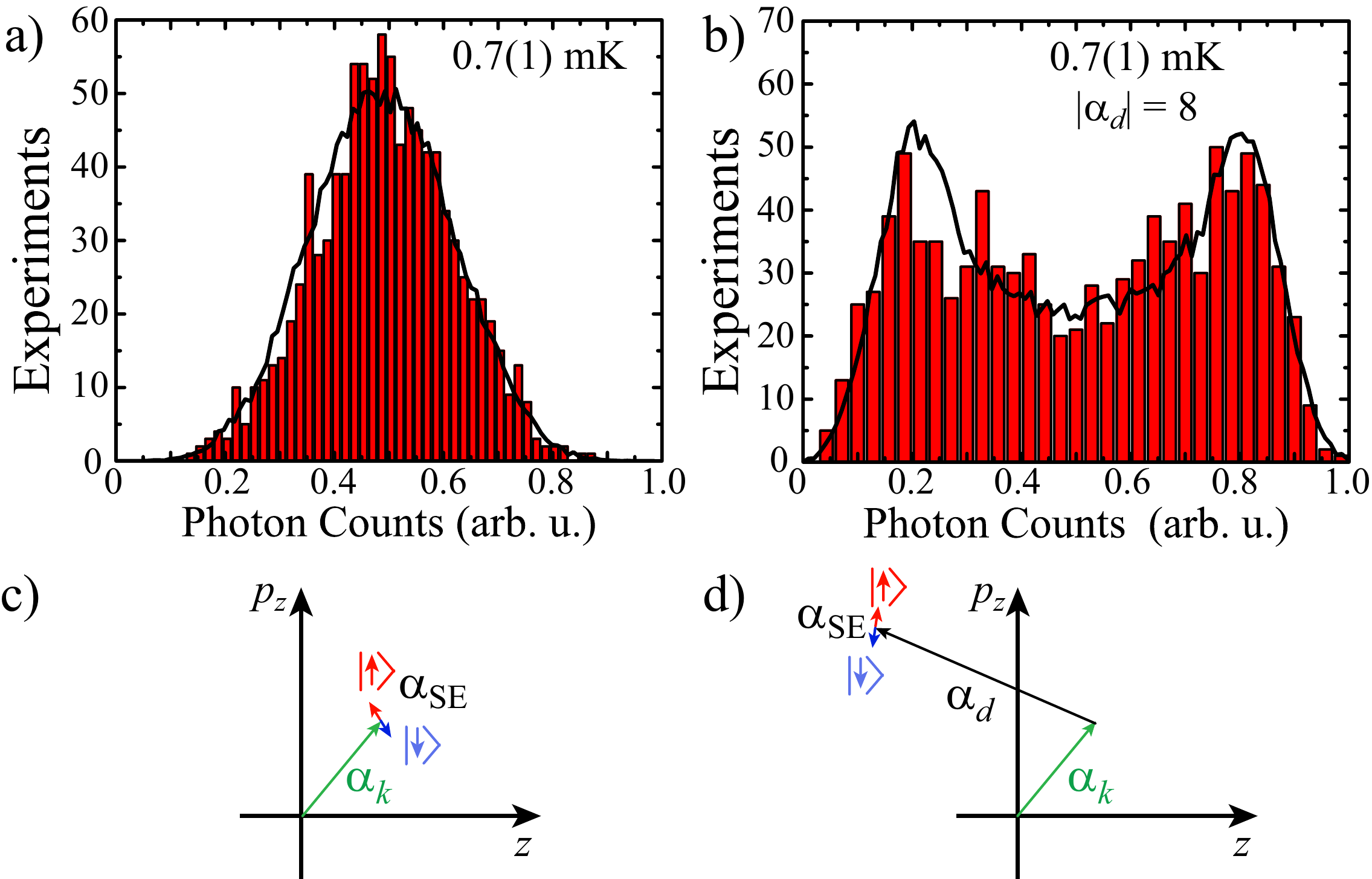}}
\caption{\label{fig3}(color online) (a) Histogram of 1000 experimental runs (red bars) under similar conditions as that shown in Fig.~\ref{fig2}b with a measured $\bar{n}=18(3)$. These data were taken without an initial coherent drive of the COM motion. Solid black histograms are obtained from Monte Carlo simulations of 50,000 experiment sequences with no adjustable parameters. (b) Measured and simulated histograms resulting from insertion of a spin-independent coherent excitation ($|\alpha_d| = 8$) before the spin dephasing measurement. The effect of coherent excitation is visible despite the fact that the relative phases of $\alpha_d$ and the ODF $\alpha_{\text{SE}}$ are uncontrolled from one experiment to another. (c, d) Example phase space trajectories of the COM mode in a frame rotating at $\omega_z$ for each Monte Carlo simulation step, $k$, both without and with the coherent RF drive. The initial displacement, $\alpha_k$, for each simulation run is chosen randomly according to Boltzmann statistics to reflect thermal fluctuations of COM motion.}
\end{figure}

We use resonant RF excitation of the COM mode to produce coherent states of motion whose square magnitude, $|\alpha_d|^2$, is larger than the thermal magnitude given by $\bar{n}$. To this end, we apply an oscillating voltage with a frequency of $\omega_z$ to the upper endcap electrode of the Penning trap for a period of 20 $\mu$s as in Ref.~\cite{Biercuk10}. This homogeneous COM mode excitation is applied following the initial Doppler cooling and state preparation pulses but before the experimental pulse sequence of Fig.~\ref{fig2}b. The amplitude of the applied voltage is 110 $\mu$V (1.2 mV m$^{-1}$ at the ion position), corresponding to a coherent excitation of $|\alpha_d|\sim 8$ ($\sim 30$ nm amplitude). Histograms compiled from 1000 experimental runs without and with the resonant RF excitation are shown in Figs.~\ref{fig3}a and~\ref{fig3}b, respectively. The absolute COM temperature is determined to be 0.7(1) mK for the sequence without coherent excitation. This elevated COM temperature is due to the connection of the direct digital synthesizer used to apply resonant RF to the upper endcap. Nevertheless, we achieve $|\alpha_d|^2/\bar{n} \sim 3.5$ for these experiments. Figure~\ref{fig3}b shows the effect of applying the coherent motional excitation, namely that the spin distribution is split into two peaks whose separation is determined by the combined coherent and ODF displacements. If the relative phase of $\alpha_d$ and $\alpha_{\text{SE}}$ were fixed for every experimental sequence, then the mean of the histogram of Fig.~\ref{fig3}a would simply be shifted to a new position corresponding to a coherent Bloch vector rotation, with the standard deviation still reflecting the COM temperature of 0.7(1) mK. However, the randomness of the relative phases of $\alpha_d$ and $\alpha_{\text{SE}}$ in combination with their constant amplitudes leads to a characteristic splitting of the spin distribution about the mean.

As in Sec.~\ref{thermalsection}, we use a Monte Carlo algorithm to simulate the experimental histograms of Fig.~\ref{fig3}a and~\ref{fig3}b. We simulate 50,000 runs to minimize noise relative to the experimental data. The simulation results given in Fig.~\ref{fig3}a (black-line histogram) use the same procedure as described in Sec.~\ref{thermalsection}. To include the coherent excitation for Fig.~\ref{fig3}b, we simply redefine the initial coherent state amplitude to be $\left(|\alpha_k|e^{i\phi_k} + |\alpha_d|e^{i\phi'_k}\right)$. Only the variables with the $k$ subscript change with each simulation run, as $|\alpha_d|=8$ is fixed for all pulse sequences. The amplitudes $|\alpha_k|^2$ once again follow a Boltzmann distribution, but here we use $T=0.7$ mK to reflect the elevated COM temperature for this set of measurements. The phases $\phi_k$ and $\phi'_k$ are uncorrelated and chosen randomly from the set $[0,2\pi)$ for each simulation run. Example paths of COM motion through phase space for each Monte Carlo simulation run are shown schematically in Figs.~\ref{fig3}c and~\ref{fig3}d both without and with the initial RF pulse, respectively. The effect of the spin-dependent ODF is depicted as two oppositely-oriented vectors leading to a separation of the spin states by $2\alpha_{\text{SE}} \sim 0.1$, while all other excitations are common to both spins.

\subsection{Heating Rates}
Motional heating in RF ion traps has gained increased prominence in recent years due primarily to studies of so-called ``anomalous heating" from ion-surface proximity~\cite{Deslauriers06,Labaziewicz08,Labaziewicz08b,Bible,Turchette00,Wang10}. Recent work suggests surface contamination as the culprit~\cite{Allcock11,Hite12}. Additionally, micromotion in RF traps limits the useable size of Coulomb crystals for quantum information and quantum simulation experiments. Penning traps use static potentials for ion confinement and therefore do not induce micromotion, enabling the formation of large ion crystals~\cite{Tan95,Itano98}. Furthermore, the large physical size of typical Penning traps means a likely insensitivity to anomalous heating processes. Despite these encouraging features, no measurements have yet been reported for ambient heating of a resolved motional mode of an ion crystal in a Penning trap. A previous study of global (not mode-resolved) ambient heating of large 3D crystals in our Penning trap estimates that background gas collisions are a primary contributor~\cite{Jensen04}. Heating rates of $\sim$65 mK/s were measured for background pressures of $\sim 4 \times 10^{-9}$ Pa ($3 \times 10^{-11}$ Torr), which translates to $\frac{d\bar{n}}{dt}\sim1.7\times10^3$ s$^{-1}$ at our trap frequency of 795 kHz.

\begin{figure}[t!]
\resizebox{8.7cm}{!}{
\includegraphics{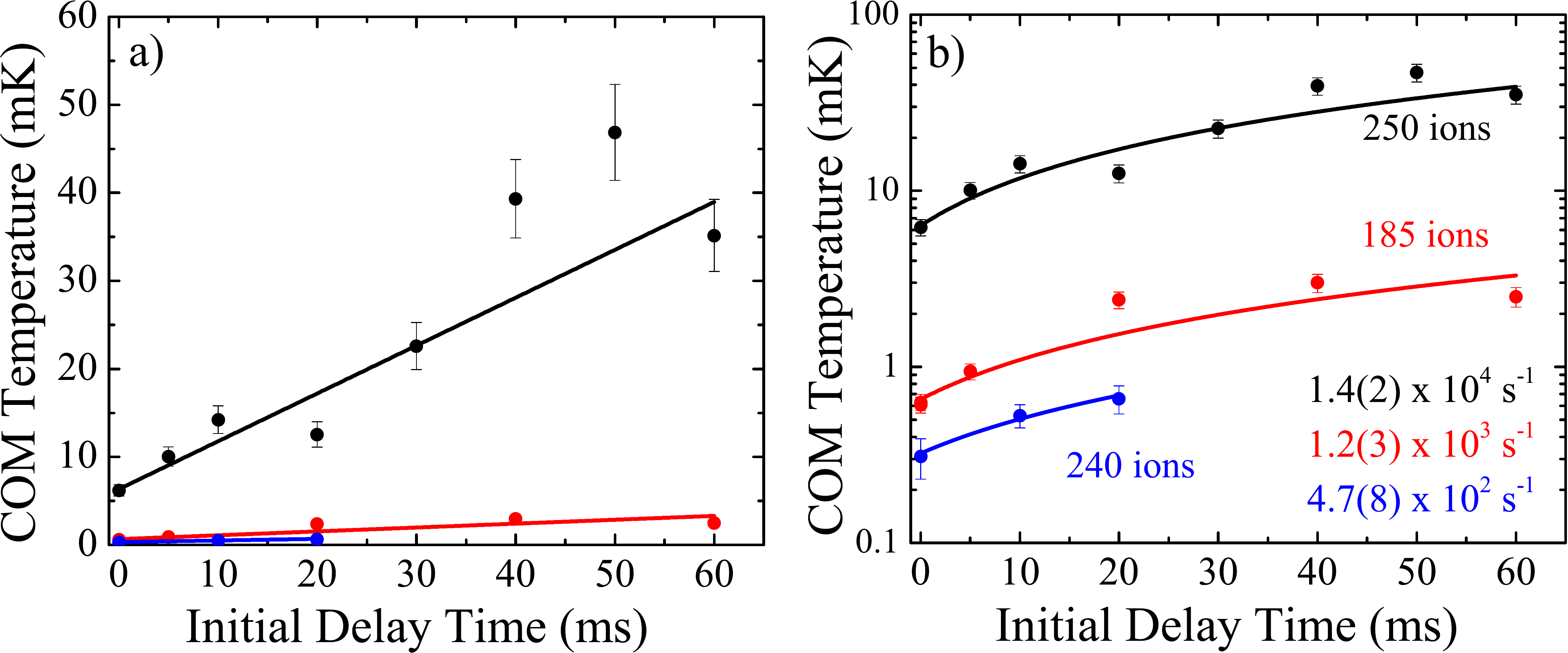}}
\caption{\label{heating}(color online) Measured COM heating rates (points with error bars) and corresponding linear fits (solid lines) plotted with both (a) linear and (b) logarithmic vertical scales. The highest heating rate of $\frac{d\bar{n}}{dt}=1.4(2) \times 10^4$ s$^{-1}$ is obtained with the trap endcap electrodes grounded through the high voltage power supplies used for initial ion loading. The two lower heating rates are measured with the endcaps shorted directly to the trap vacuum ground at the high voltage vacuum feedthrough. The variation between the two lowest rates is representative of the range of heating rates measured thus far under the given conditions.}
\end{figure}

We apply the thermometry techniques summarized in Fig.~\ref{fig2} to obtain an initial measurement of the ambient heating rate of the axial COM mode of our 2D crystals. We include a variable delay between initial Doppler cooling/state preparation and application of the experimental pulse sequences. Any increase in $\bar{n}$ over this initial delay period is measured in the subsequent decoherence or dephasing measurement. Figure~\ref{heating} shows measured absolute COM temperatures as a function of initial delay (points with error bars) along with linear fits to each data set (solid lines). The data of panels~\ref{heating}a and~\ref{heating}b are identical, but plotted on linear and logarithmic vertical axes, respectively, for clarity. The fitted slopes reflecting $\frac{d\bar{n}}{dt}$ for each curve are displayed in Fig~\ref{heating}b and color-coded to match the corresponding data set.

We measure the largest heating rate of $1.4(2)\times 10^{4}$ s$^{-1}$ (black points) when the trap endcap electrodes are held at 0 V using the high voltage power supplies responsible for initial ion loading and transport. Small voltage fluctuations from these power supplies as well as electromagnetic interference along the connecting cables are the likely cause of this COM heating. Upon grounding the endcaps directly to the vacuum chamber at the high-voltage feedthrough, we observe an order-of-magnitude drop in the heating rate to between $4.7(8) \times 10^2$ s$^{-1}$ (blue points) and $1.2(3) \times 10^3$ s$^{-1}$ (red points). This approximate factor-of-two variation in heating rates is representative of our day-to-day observations with different ion samples, and we see no evidence of a correlation between total ion number and heating rate with the endcap electrodes grounded at the feedthrough. The mean of the two lowest heating rates corresponds to $\sim$30 mK/s, which is near previous collisional heating estimates~\cite{Jensen04}. Additional filtering of the central ring and rotating-wall electrodes outside the vacuum envelope yielded no measurable improvement in COM heating~\footnote{Heating rate measurements of single ions in small RF traps are generally thought to be insensitive to background gas collisions. However, we believe our ambient heating measurements on hundreds of ions in a deep Penning trap should be sensitive to background gas collisions.}.

Because the distance from the trapped-ion arrays to the trap electrode surfaces ($\geq$2 cm) is large compared to the diameter of the planar array ($<$0.5 cm), electric field noise from trap electrode surfaces will be uniform across the array and preferentially heat the COM mode. For an array with $N$ ions this results in a linear dependence of the COM heating rate on ion number due to uniform electric field noise~\cite{HomeAMOAdvances13}. The lack of an observed $N$-dependence in the measured ambient heating rate indicates the source of the heating is likely not electric field noise.  However, any potential anomalous heating must be less than the measured $\sim$10$^3$ s$^{-1}$ ambient heating rate.  Dividing this limit by the number of trapped ions ($N \sim 200$) gives a limit on the anomalous heating rate for a single trapped ion of $\sim$5 s$^{-1}$ at a trap frequency of 795 kHz.

\section{Conclusion}
In summary, we have demonstrated a new technique for analyzing the motional state of a resolved ion crystal mode. The methods presented here do not rely on stimulated Raman transitions or Doppler linewidth analysis, and are in principle applicable to any resolved motional mode at any temperature provided the Lamb-Dicke confinement criterion is satisfied for the given mode. The sensitivity of our spin dephasing measurements allows for a regime of operation with negligible spin-motion entanglement and spontaneous emission decoherence. Monte Carlo simulations based on the semiclassical description of Sec.~\ref{Theory} are in excellent agreement with spin dephasing measurements. Furthermore, we observe a clear distinction between coherent and thermal states of motion, despite the randomness of the RF drive phase relative to our optical dipole force.

The methods and analysis presented here enable very sensitive detection of coherently driven motion of a trapped-ion crystal, and may be used to phase-sensitively detect weak forces~\cite{Biercuk10,Schreppler13}. Section III can be used to estimate and optimize the force detection sensitivity for a given $N$ and temperature of the trapped-ion crystal. We estimate that the spin-motion coupling technique discussed here could improve on the force sensitivity obtained in Ref.~\cite{Biercuk10} by more than an order-of-magnitude.

We also present the first measurements of ambient heating of a resolved mode of motion in a Penning trap. Future crystal heating measurements will include other resolved transverse motional modes with the goal of more clearly distinguishing between electric field fluctuations (mode-specific, $N$-dependent) and background gas collisions (mode- and $N$-independent). Nevertheless, we demonstrate that low heating rates are indeed achievable in Penning ion traps.

\begin{acknowledgments}
This work was supported by the DARPA-OLE program and NIST. The authors thank J. P. Home, K. R. A. Hazzard, M. Foss-Feig, and A. M. Rey for useful discussions as well as S. Kotler and J. P. Gaebler for comments on the manuscript. This manuscript is a contribution of NIST and not subject to U.S. copyright.
\end{acknowledgments}

\appendix
\section{Calculating Bloch Vector Length using Fock States} \label{AppA}
A thermal motional state is described by a density matrix that is a statistical mixture of Fock states. This motivates a calculation that assumes each experiment begins with the system in the state, $|\!\!\uparrow\rangle|n\rangle$, where $|n\rangle$ is the harmonic oscillator Fock state of COM motion. The first $\pi/2$ pulse of the Ramsey sequence yields the qubit rotation
\begin{equation}
|\psi_1\rangle=\hat{R}(\frac{\pi}{2},0)|\!\!\uparrow\rangle|n\rangle = \frac{1}{\sqrt{2}}\left(\upstate + \downstate \right)|n\rangle.
\end{equation}
The spin-dependent ODF then produces displaced Fock states, $|\alpha,n\rangle$, as~\cite{Wunsche91}
\begin{equation}
|\psi_2\rangle = \hat{D}_{\text{SD}}(\alpha)|\psi_1\rangle = \frac{1}{\sqrt{2}}\left( \upstate |\alpha,n\rangle
+\downstate |-\alpha,n\rangle \right).
\end{equation}
Note that $|\psi_2\rangle$ involves entanglement of spin and motional degrees of freedom for nonzero $\alpha$. We now apply the final $\pi/2$ pulse whose phase is identical to the first ($\Delta\phi=0$) to obtain
\begin{eqnarray}\label{psif}
|\psi_f\rangle &=& \frac{1}{2}
\upstate\left( |\alpha,n\rangle - |-\alpha,n\rangle \right) \nonumber \\
  && +\frac{1}{2}\downstate \left( |\alpha,n\rangle + |-\alpha,n\rangle \right).
\end{eqnarray}
The probability of measuring $\upstate$ for state $|\psi_f\rangle$ depends on the overlap of $|\alpha, n\rangle$ and $|-\alpha, n\rangle$, and is given by
\begin{equation}
P_{\uparrow}^{(n)} = \frac{1}{2}\left(1 - L_n\left(4|\alpha|^2\right)e^{-2|\alpha|^2}\right)
\end{equation}
where $L_n$ is the Laguerre polynomial of order $n$. Our fluorescence detection is insensitive to the ion motional state, so we perform a Boltzmann-weighted thermal average over all Fock states to obtain
\begin{eqnarray}
P_{\uparrow} &\equiv& \langle P_{\uparrow}^{(n)}\rangle_{th} \nonumber \\
&=& \frac{1}{2}\left(1 - e^{-2|\alpha|^2}\langle L_n\left(4|\alpha|^2\right)\rangle_{th}\right) \nonumber \\
&=& \frac{1}{2} \left(1 - e^{-2|\alpha|^2(2\bar{n}+1)}\right). \label{PupFock}
\end{eqnarray}
In the above equation, $\bar{n}=\left(e^{\beta}-1\right)^{-1}$ is the average COM mode occupation number for a thermal state at temperature, $T$, and $\beta = \hbar \omega_z / k_B T$. In the absence of the spin-dependent displacement, $P_{\uparrow}=0$ for this pulse sequence. However, as the displacement amplitude increases, $P_{\uparrow}$ takes on positive values that increase with mode occupation -- eventually saturating at $P_{\uparrow}=0.5$ corresponding to complete loss of spin coherence. In Ref.~\cite{Sawyer12}, we describe using this decoherence signature to perform mode spectroscopy and thermometry on a planar array of ions.

We may also calculate the expectation value for the $N$-spin composite Bloch vector, $\langle \hat{S}_z \rangle=\langle \sum_{i=1}^N \frac{\hat{\sigma}^z_i}{2}\rangle = \frac{N}{2}\langle \hat{\sigma}^z \rangle$ using the final state of Eq.~\ref{psif}. We obtain
\begin{eqnarray}
\langle \hat{S}_z \rangle &=& -\frac{N}{2}L_n\left(4|\alpha|^2\right)e^{-2|\alpha|^2} \\
\langle \langle \hat{S}_z \rangle\rangle_{th} &=& -\frac{N}{2}e^{-2|\alpha|^2(2\bar{n}+1)}.
\end{eqnarray}

We note that the calculation in this Appendix, which uses the Fock state basis, motivates a picture of spin decoherence produced by entanglement of the spin and motional degrees of freedom.  However, in the manuscript we show that for coherent input states of motion, spin decoherence can be explained by dephasing without resorting to quantum entanglement of spin and motion.

\section{Calculating Dephasing using Fock States} \label{AppB}
If the phase of the final microwave $\pi/2$-pulse is shifted by $\pi/2$ (e.g. $\hat{R}(\frac{\pi}{2},\frac{\pi}{2})$, $\Delta\phi = \pi/2$), then the composite Bloch vector will remain in the equatorial plane of the Bloch sphere and $\langle \hat{S}_z \rangle = \langle \hat{S}_z \rangle_{th}=0$. Fock states therefore produce no coherent spin rotation due to the spin-dependent ODF.

To calculate dephasing, we must compute pairwise spin correlations of the form $\langle \hat{\sigma}^z_1 \hat{\sigma}^z_2 \rangle$ as shown in Eq.~\ref{2ndmom}. As in Appendix~\ref{AppA}, we will consider the initial state of COM motion to be a Fock state, $| n \rangle$. In contrast with the Bloch vector length calculation that requires only a single spin, we here consider a two-spin system whose full initial state is $|\!\! \uparrow \uparrow\rangle |n\rangle$. We construct the necessary two-spin rotation matrices, $\hat{R}^{(2)}(\frac{\pi}{2},0)$ and $\hat{R}^{(2)}(\frac{\pi}{2},\frac{\pi}{2})$, using Kronecker products as follows:

\begin{small}
\begin{eqnarray}
\hat{R}^{(2)}(\frac{\pi}{2},0) &\equiv& \hat{R}(\frac{\pi}{2},0) \otimes \hat{R}(\frac{\pi}{2},0) \nonumber \\
&=& \frac{1}{2}
\left(
\begin{array}{rrrr}
   1 & -1 & -1 & 1 \\
   1 & 1 & -1 & -1  \\
   1 & -1 & 1 & -1 \\
   1 & 1 & 1 & 1 \\
\end{array}
\right) \\
\hat{R}^{(2)}(\frac{\pi}{2},\frac{\pi}{2}) &\equiv& \hat{R}(\frac{\pi}{2},\frac{\pi}{2}) \otimes \hat{R}(\frac{\pi}{2},\frac{\pi}{2}) \nonumber \\
&=& \frac{1}{2}
\left(
\begin{array}{rrrr}
   1 & i & i & -1 \\
   i & 1 & -1 & i  \\
   i & -1 & 1 & i \\
   -1 & i & i & 1
\end{array}
\right).
\end{eqnarray}
\end{small}
We can now define the final state, $|\psi_f\rangle$, after the pulse sequence of Fig.~\ref{ramtheory} ($\Delta \phi = \frac{\pi}{2}$) as
\begin{equation}
|\psi_f\rangle = \hat{R}^{(2)}(\frac{\pi}{2},\frac{\pi}{2})  \hat{D}_{\text{SD}}(\alpha) \hat{R}^{(2)}(\frac{\pi}{2},0) |\!\! \uparrow\uparrow\rangle|n\rangle,
\end{equation}
where $\hat{D}_{\text{SD}}=\exp\left(\left[\alpha\hat{a}^{\dagger}-\alpha^{*}\hat{a}\right]\sum_{i=1}^{2}\hat{\sigma}_{i}^{z}\right)$ and $\alpha$ is the spin-dependent displacement amplitude. The expectation value $\langle \hat{\sigma}^z_1 \hat{\sigma}^z_2 \rangle = \langle\psi_f|\hat{\sigma}^z_1 \hat{\sigma}^z_2|\psi_f\rangle$ and corresponding thermal average are
\begin{eqnarray}
\langle \hat{\sigma}^z_1 \hat{\sigma}^z_2 \rangle &=& \frac{1}{2} \left[ 1 - L_n(16|\alpha|^2)e^{-8|\alpha|^2} \right] \\
\langle\langle \hat{\sigma}^z_1 \hat{\sigma}^z_2 \rangle \rangle_{th} &=& \frac{1}{2} \left[ 1 - e^{-8|\alpha|^2(2\bar{n}+1)} \right]. \label{correlator_fock}
\end{eqnarray}
As is the case with the Bloch vector length calculation, Eq.~\ref{correlator_fock} agrees with our earlier result using coherent states (Eq.~\ref{2ndmomcohshort}) in the limit $\bar{n} \sim \beta^{-1}$.

\end{document}